\documentclass[aps,prl,a4paper,superscriptaddress,twocolumn,showpacs,preprintnumbers,amsmath,amssymb]{revtex4-1}

\usepackage{graphicx}
\usepackage{dcolumn}
\usepackage{bm}
\usepackage{amsmath}
\usepackage[colorlinks=true, linkcolor=black, citecolor=black, urlcolor=blue]{hyperref}
\usepackage{dsfont}
\usepackage{color}
\usepackage[mathscr]{euscript}
\usepackage{soul, xcolor}
\setstcolor{red}

\definecolor{mygreen}{rgb}{0,0.5,0} 
\definecolor{myblue}{rgb}{0,0,0.75} 
\definecolor{mymagenta}{cmyk}{0,1,0,0.12} 

\newcommand{\gtext}{}

\begin{document}


\title{Spontaneous $\mathcal{PT}$ symmetry breaking of a ferromagnetic superfluid in a gradient field}

\newcommand{\ICFOAddress}{ICFO -- Institut de Ciencies Fotoniques, The Barcelona Institute of Science and Technology, 08860 Castelldefels (Barcelona), Spain}
\newcommand{\ICREAAddress}{ICREA -- Instituci\'{o} Catalana de Re{c}erca i Estudis Avan\c{c}ats, 08015 Barcelona, Spain}

\author{T. Vanderbruggen}
\author{S.~Palacios~{\`Alvarez}}
\author{S.~Coop}
\author{N.~Martinez de Escobar}
\affiliation{\ICFOAddress}

\author{M. W. Mitchell}
\affiliation{\ICFOAddress}
\affiliation{\ICREAAddress}

\date{\today}

\begin{abstract}
We consider the interaction of a ferromagnetic spinor Bose-Einstein condensate with a magnetic field gradient. The magnetic field gradient realizes a spin-position coupling that explicitly breaks time-reversal symmetry $\mathcal{T}$ and space parity $\mathcal{P}$, but preserves the combined $\mathcal{PT}$ symmetry. We observe using numerical simulations, a phase transition spontaneously breaking this remaining symmetry. The transition to a low-gradient phase, in which gradient effects are frozen out by the ferromagnetic interaction, suggests the possibility of high-coherence magnetic sensors unaffected by gradient dephasing.
\end{abstract}


\maketitle


\noindent \textit{Introduction} -- The discovery of the very complex vacuum in superfluid $^3$He proved to be highly stimulating to the theory of symmetry breaking \cite{bruder1986} and topological defects \cite{mermin1979,salomaa1987}. New features of quantum gases emerged with the \gtext{first realizations of} spinor Bose-Einstein condensates (SBEC) \cite{Stenger1998,barrett2001}, thanks to the many degrees of freedom -- both internal and external -- and to the excellent control of the experimental parameters. SBECs are extremely rich and versatile systems to study complex quantum vacuua \cite{stamper-kurn2013}, for example to test the validity of universal phenomena like the Kibble-Zurek mechanism \cite{damski2007,swislocki2013,Navon2015}, or to study Goldstone modes such as gapless magnons \cite{marti2014}.

The coupling of SBECs to magnetic fields has been exploited to study quantum phase transitions in SBECs \cite{sadler2006, black2007, jacob2012}, and to realize point-like topological defects such as Dirac monopoles \cite{ray2014, pietila2009, PhysRevA.84.063627} and 2D skyrmions \cite{choi2012}. In these works, spin symmetries and topology were  induced by strong  gradients, e.g. $37$~mT/m in Ref.~\cite{ray2014}. Here we show that via a quantum phase transition at lower gradients $\sim 0.5$~mT/m, the ferromagnetic interaction can ``freeze out'' the gradient effect. This suggests the possibility of magnetic sensors free from gradient dephasing, a practical limitation in coherent magnetometry \cite{BudkerNP2007,VengalattorePRL2007, ShahNP2007,KoschorreckAPL2011,SmithJPB2011,SewellPRL2012, BehboodAPL2013}.
  
We use group theoretical methods that have proven fruitful in the analysis and classification of SBEC phases \cite{bruder1986, kawaguchi2011}. The interaction with the gradient realizes an interesting spin-position coupling that explicitly breaks the parity $\mathcal{P}$ and time-reversal $\mathcal{T}$ symmetries, only preserving the combined $\mathcal{PT}$ symmetry. Numerically solving the Gross-Pitaevskii equations of the system, we observe that below a critical value of the magnetic field gradient, the $\mathcal{PT}$ symmetry is spontaneously broken and a nonzero overall magnetization appears. This occurs when the ferromagnetic interactions dominate the coupling energy between the gradient field and the spins, resulting in a globally polarized condensate. Moreover, we  observe that this effect is associated with a phase transition. Interestingly, discrete, and in particular $\mathcal{PT}$, symmetry breaking is also observed in Bose gases with spin-orbit coupling \cite{xu2012,hu2012,sinha2011}.

\noindent \textit{System and mean-field energy} -- We consider a spin-1 BEC with ferromagnetic interactions in the presence of a magnetic field gradient. More specifically, the numerical calculations are performed for a $^{87}$Rb SBEC in the $F=1$ ground state, which is composed of three Zeeman sublevels $m_{F} = -, \, 0, \, +$.

Within the mean-field approximation, the spin-1 BEC is described by a spinorial field with three complex components $\Psi(\mathbf{r}) \equiv \left[ \psi_{-}(\mathbf{r}), \psi_{0}(\mathbf{r}), \psi_{+}(\mathbf{r}) \right]^{T}$, where $\mathbf{r} = (x,y,z)^{T}$ are the spatial coordinates, and $\psi_{\mu}(\mathbf{r})$ is the mean-field wavefunction for the atomic distribution in the magnetic sublevel $m_{F} = \mu$. 

The mean-field energy density of the system, coupled to a magnetic field distribution $\mathbf{B}(\mathbf{r})$, is \cite{doi:10.1143/JPSJ.67.1822, ho1998}
\begin{align}
\mathcal{E}[\Psi] =& \, \psi^{*}_{\alpha} \left[ -\frac{\hbar^{2}}{2m} \nabla^{2} + V \right] \psi_{\alpha}  + \frac{c_{0}}{2} \, \left( \psi^{*}_{\alpha} \psi_{\alpha} \right)^{2} \nonumber \\
& \, - g \mu_{B} B_{i} \psi^{*}_{\alpha} F_{\alpha \beta}^{i} \psi_{\beta} + \frac{c_{2}}{2} \, \psi^{*}_{\alpha} \psi^{*}_{\mu} F_{\alpha \beta}^{i} F_{\mu \nu}^{i} \psi_{\beta} \psi_{\nu},
\label{eq:MF_density}
\end{align}
where the Latin letters designate the spatial coordinates ($x,y,z$) and the Greek letters the spin coordinates ($-,0,+$). The first term is the sum of the kinetic energy ($m$ is the atom's mass) and of the trapping potential $V(r)$, assumed to be harmonic and spatially isotropic. The third term is the energy density resulting from the coupling with the magnetic field $B_{i}(\mathbf{r})$, where $g$ is the gyromagnetic ratio, $\mu_{B}$ the Bohr magneton, and $F^{i}$ is the generator of spin-1 rotations around the $i$ axis. The terms containing $c_0$ and $c_2 < 0$ describe the spin-independent and ferromagnetic spin-dependent collisional energies, respectively.

The magnetic field is chosen to be a pure gradient (no bias) along the $z$ axis, i.e. the divergenceless field
\begin{equation}
\mathbf{B}(\mathbf{r}) = B'(x/2,y/2,-z)^{T} = B' \Lambda \mathbf{r},
\end{equation}
where we defined the metric $\Lambda \equiv \mathrm{diag} (1/2,1/2,-1)$. The mean-field energy density associated with the gradient coupling is thus
\begin{equation}
\mathcal{E}_{G} [\Psi](\mathbf{r}) = -g \mu_{B} B' \, \Lambda \mathbf{r} \cdot \mathbf{F}(\mathbf{r}),
\end{equation}
where $\mathbf{F} \equiv (\mathscr{F}_{x}, \mathscr{F}_{y}, \mathscr{F}_{z})^{T}$ and $\mathscr{F}_{i} \equiv \psi^{*}_{\alpha} F_{\alpha \beta}^{i} \psi_{\beta}$. This interaction realizes a spin-position coupling (similar to spin-orbit coupling $\mathbf{L} \cdot \mathbf{F}$). The total mean field energy related to the interaction with the magnetic field gradient is 
\begin{equation}
E_{G} [\Psi] \propto \int \left[ -\frac{1}{2} x \, \mathscr{F}_{x}(\mathbf{r}) -\frac{1}{2} y \, \mathscr{F}_{y}(\mathbf{r}) + z \, \mathscr{F}_{z}(\mathbf{r}) \right] d^{3}r.
\label{eq:MF_gradient}
\end{equation}

\noindent \textit{Symmetries} -- We now study the symmetries of the problem, that is, the transformations that leave invariant the mean-field energy $E[\Psi] = \int \mathcal{E}[\Psi] (\mathbf{r}) \, d^{3}r$. Focussing on the invariance of the gradient coupling energy [Eq.~(\ref{eq:MF_gradient})], we observe that the spin-position coupling explicitly breaks several symmetries, both continuous and discrete.

We first consider the continuous symmetries. In absence of magnetic field gradient, the energy is invariant under both spin and space rotations. When a gradient is present, the system is only invariant under combined spin-space rotations around the $z$ axis. More precisely, we define the operator $R_{z} (\theta)$ acting on $\Psi$ as
\begin{equation}
R_{z} (\theta) \Psi (\mathbf{r}) = e^{i \theta F^{z}} \Psi \left( e^{- i \theta L^{z}} \mathbf{r} \right),
\end{equation}
where $L^{z}$ is the generator of spatial rotations around the $z$ axis. The set of all such transformations for $\theta \in \mathds{R}$ is a one parameter rotation group, denoted as SO(2)$_{F^{z} + L^{z}}$. From Eq.~(\ref{eq:MF_gradient}), it is straightforward to show that
\begin{equation}
E_{G} [R_{z} (\theta) \Psi] \propto \int \Big[ -\frac{x'}{2} \, \mathscr{F}_{x}\left( \mathbf{r}' \right) -\frac{y'}{2} \, \mathscr{F}_{y}\left( \mathbf{r}' \right) + z' \, \mathscr{F}_{z}\left( \mathbf{r}' \right) \Big] d^{3}r,
\end{equation}
where $\mathbf{r}' \equiv e^{- i \theta L^{z}} \mathbf{r}$. After the change of variable $\mathbf{r}' \rightarrow \mathbf{r}$, we obtain $E_{G} [R_{z} (\theta) \Psi] = E_{G} [\Psi]$, proving the invariance under SO(2)$_{F^{z} + L^{z}}$. In other words, the spin-position coupling explicitly breaks the  $\mathrm{SO(3)}_{\mathbf{F}} \times \mathrm{SO(3)}_{\mathbf{L}}$ symmetry into $\mathrm{SO(2)}_{F^{z} + L^{z}}$. Therefore, the magnetization $\mathbf{F}$ and the orbital angular momentum $\mathbf{L}$ are no longer independently conserved, and only the total longitudinal angular momentum $F^{z} + L^{z}$ is conserved \cite{lesanovsky2005}.

Beyond continuous symmetries, the mean-field energy also exhibits discrete symmetries. We define the spatial inversions, corresponding to the parity symmetry, as 
\begin{equation}
\mathsf{P}_{x} : \Psi(x,y,z) \mapsto \Psi(-x,y,z),
\end{equation}
and similarly for $\mathsf{P}_{y}$ and $\mathsf{P}_{z}$. The spin inversions, corresponding to the time-reversal symmetry, are defined as
\begin{align}
\mathsf{T}_{x} &: \psi_{\mu} \mapsto (-1)^{\mu} \psi_{\mu}^{*} ,\\
\mathsf{T}_{y} &: \psi_{\mu} \mapsto \psi_{\mu}^{*}, \\
\mathsf{T}_{z} &: \psi_{\mu} \mapsto \psi_{-\mu}^{*}.
\end{align}
The energy functional $E_{G}$ is not invariant under parity nor time-reversal\footnote{\gtext{We consider the system composed of the spins, where an externally-imposed magnetic field breaks time symmetry. This is the same convention used for example in the description of non-reciprocal devices, such as an optical isolator.}}, however it is invariant under a combined space-spin inversion. For example, we consider the combined action of $\mathsf{P}_{x}$ and $\mathsf{T}_{x}$ on the mean-field energy
\begin{align}
E_{G}[\mathsf{P}_{x}\mathsf{T}_{x}\Psi] &\propto \int \Big[ +\frac{1}{2} x \, \mathscr{F}_{x}(-x,y,z) -\frac{1}{2} y \, \mathscr{F}_{y}(-x,y,z) \nonumber \\
& \;\;\;\;\;\;\;\;\;\;\;\;\;\;\;\;\; + z \, \mathscr{F}_{z}(-x,y,z) \Big] d^{3}r,
\end{align}
and a change of variable $x \rightarrow -x$ leads to $E_{G}[\mathsf{P}_{x}\mathsf{T}_{x}\Psi] = E_{G}[\Psi]$. Therefore, $E_{G}$ is invariant under the discrete group $\mathcal{PT}_{x} \equiv \left\{ \mathds{1}, \mathsf{P}_{x}\mathsf{T}_{x} \right\}$. By analogous arguments, the mean-field energy is also invariant under $\mathcal{PT}_{y}$ and $\mathcal{PT}_{z}$, defined analogously. Therefore, $E_{G}$ is fully symmetric under $\mathcal{PT} \equiv \mathcal{PT}_{x} \times \mathcal{PT}_{y} \times \mathcal{PT}_{z}$ \footnote{Note that $\mathcal{PT}_{y} \subset \mathrm{SO(2)}_{F^{z} + L^{z}} \times \mathcal{PT}_{x}$, since $\mathsf{P}_{y}\mathsf{T}_{y} = R_{z} (\pi) \cdot \mathsf{P}_{x}\mathsf{T}_{x}$. Thus, given the invariance of the system under $\mathrm{SO(2)}_{F^{z} + L^{z}}$, the symmetry $\mathcal{PT}_{x/y} \times \mathcal{PT}_{z}$ is sufficient to achieve the full $\mathcal{PT}$ symmetry.\label{ftn:factor_group}}.

Finally, the energy is also invariant under a phase-shift $\Psi(\mathbf{r}) \rightarrow e^{i \phi} \Psi(\mathbf{r})$, i.e., under the gauge group $\mathrm{U(1)}_{\phi}$. As a consequence, the global symmetry group of the mean-field energy is
\begin{equation}
G = \mathrm{U(1)}_{\phi} \times \mathrm{SO(2)}_{F^{z} + L^{z}} \times \mathcal{PT}.
\end{equation}

\noindent \textit{Numerical simulations} -- Whereas the $\mathrm{U(1)}_{\phi}$ symmetry is spontaneously broken by the Bose transition, different breaking scenarios are possible for the remaining factors of the group $G$, as a result of the competition between magnetic field gradient coupling and spin-dependent interactions. Here, using a numerical simulation, we study the spontaneously broken symmetries and characterize the different phases.

The mean-field evolution of the system is determined by three coupled Gross-Pitaevskii equations associated to the mean-field energy density Eq.~(\ref{eq:MF_density}), explicitly
\begin{align}
i \hbar \partial_{t} \psi_{\alpha} =& \left[ -(\hbar^{2}/2m) \nabla^{2} + V(r) \right] \psi_{\alpha}  + c_{0} \psi_{\beta}^{*} \psi_{\beta} \psi_{\alpha} \nonumber  \\
& - g \mu_{B} B' \Lambda_{ij} r_{j} F_{\alpha \beta}^{i} \psi_{\beta} + c_{2}
\psi^*_\mu  F_{\mu \nu}^{i} \psi_\nu F_{\alpha \beta}^{i} \psi_{\beta},
\label{eq:sGPE}
\end{align}
where $\alpha \in \{ +,0,- \}$. The trapping potential $V(r)$ is chosen isotropic and harmonic with frequency 100~Hz. The atomic species is $^{87}$Rb in the $F=1$ hyperfine ground state, thus $m=1.44 \times 10^{-25}$~kg, $g=-1/2$, and the collisional interaction parameters are $c_{0} = 5.16 \times 10^{-51}$~J$\cdot$m$^3$ and $c_{2} = -2.39 \times 10^{-53}$~J$\cdot$m$^3$. Since $c_{2}<0$ the spins experience ferromagnetic interactions.

\begin{figure}[!h]
\begin{center}
\includegraphics[width=8.5cm,keepaspectratio]{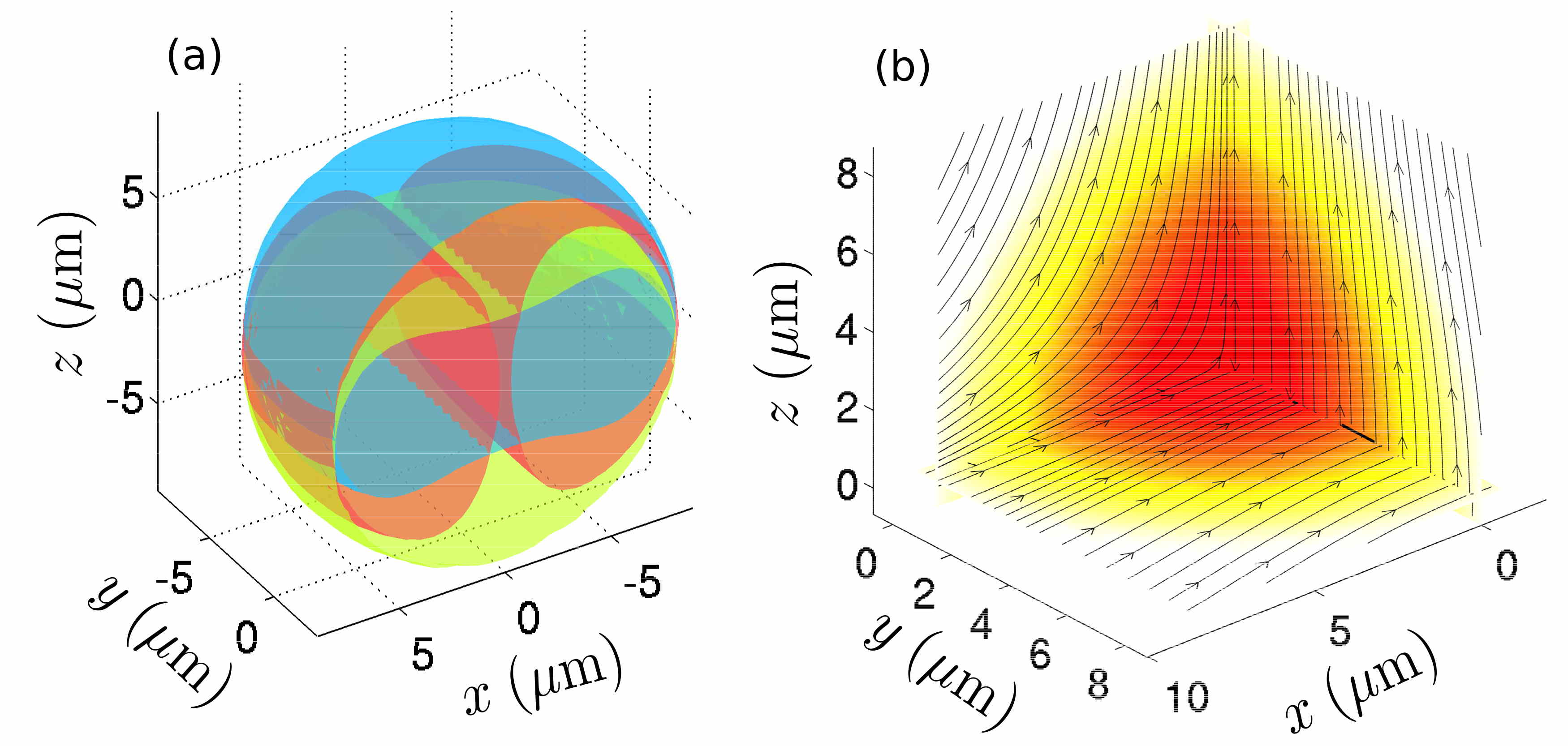}
\caption{(Color online) (a) Isoprobability surfaces of the spinor components: $\psi_{-}$ (yellow, lower), $\psi_{0}$ (red, middle) and $\psi_{+}$ (blue, upper). (b) Streamlines (curves tangent to the spinor field) in three orthogonal planes, and the color map shows the total atomic density. The magnetic field gradient is $B'=3 \times 10^{-4}$~T/m and the number of atoms is $5 \times 10^{5}$.}
\label{fig:isosurf}
\end{center}
\end{figure}

\newcommand{\ket}[1]{|#1 \rangle}
The stationary state of the GPEs is numerically determined using an imaginary-time method from the GPELab 3D solver \cite{antoine2014}. The time and space discretization is achieved by a backward Euler spectral FFT scheme. The initial state is an equal superposition of all $m_{F}$ states: $\Psi(\mathbf{r}) = \sqrt{f(\mathbf{r})} \, \left[1,1,1 \right]^{T}$, that is, a state pointing in the $+x$ direction, and where $f(\mathbf{r})$ is a Gaussian distribution with width the harmonic oscillator radius of the trapping potential $V(r)$.

As shown on a simulation result in Fig.~\ref{fig:isosurf}, the magnetic field gradient induces a spatial separation of the various $m_{F}$ components along the $z$ axis. We also observe in Fig.~\ref{fig:isosurf}~(b) that the spins in the $x-y$ plane are mainly oriented along $x$, thus the $\mathrm{SO(2)}_{F^{z} + L^{z}}$ symmetry is spontaneously broken by the ferromagnetic interaction. This effect also appears in Fig.~\ref{fig:isosurf}~(a), where the $\psi_{0}$ wavefunction is not invariant by rotation around $z$. The prevailing direction is $x$ in the current simulations due to the choice of initial conditions, however it can be any direction $\hat{\mathbf{f}}$ in the $x-y$ plane \footnote{We use $x$-polarized initial states for simplicity and speed of convergence. To check robustness, we run representative cases with $z$-oriented initial states and nearly $z$-oriented states. These converge to the same final states (modulo a $z$ rotation) as do $x$-polarized initial states}.

\begin{figure}[!h]
\begin{center}
\includegraphics[width=7.5cm,keepaspectratio]{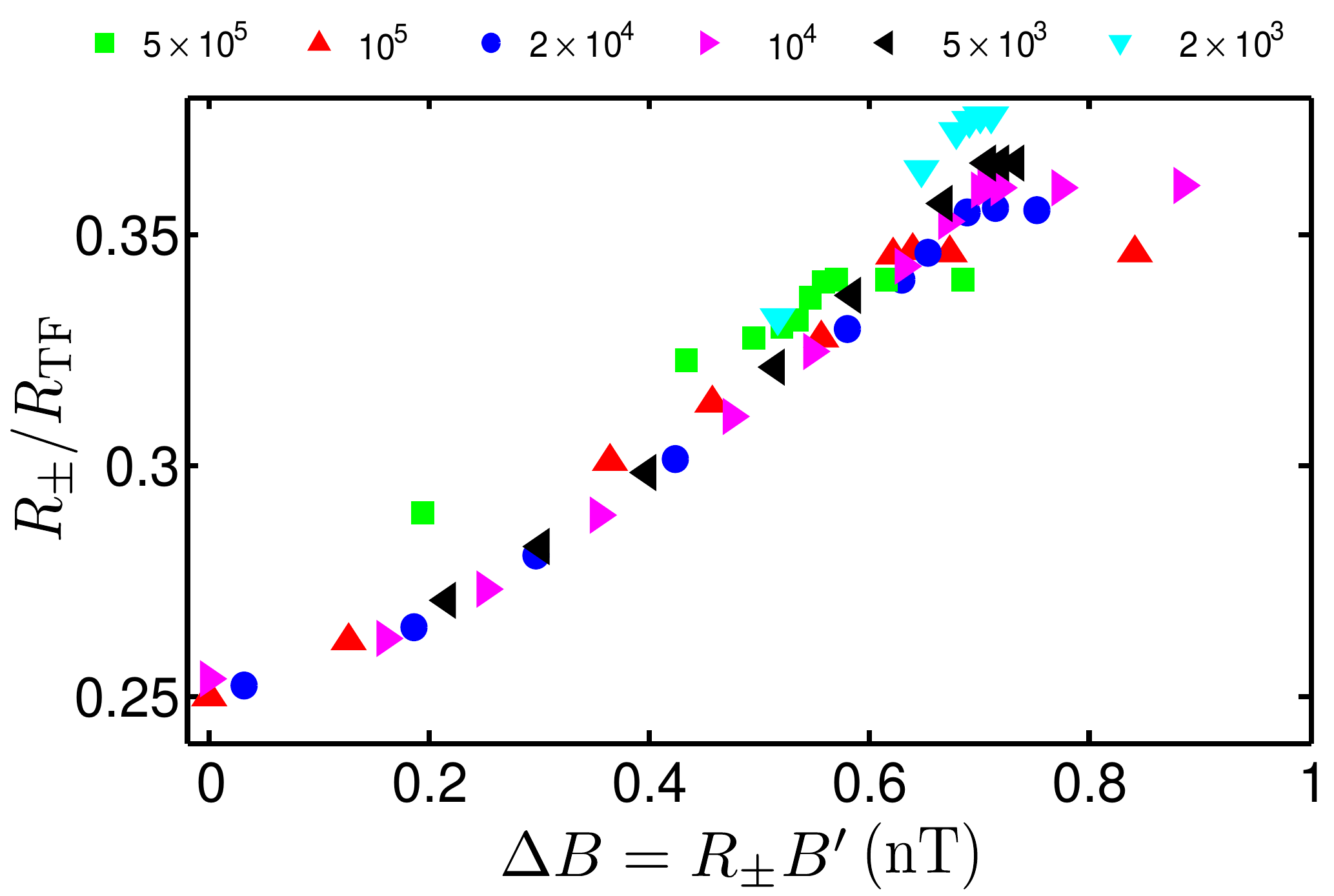}
\caption{Cloud radius over the Thomas-Fermi radius versus $B$ for various atom numbers. The legend at the top indicates the number of atoms.}
\label{fig:res_sim_radius}
\end{center}
\end{figure}

\newcommand{\Rpm}{R_{\pm}}

A natural measure of gradient strength is  $\Delta B = \Rpm B'$, where $\Rpm$ is the mean-square radius of the $\pm$ components.  As shown in Fig.~\ref{fig:res_sim_radius}, this radius, normalized by the Thomas-Fermi radius $R_{\rm TF}$, is nearly independent of the atom number in the range $2000 \lesssim N \lesssim 10^5$.  Below this range the low density voids the Thomas-Fermi approximation, and above this range the large size of the system exhibits the first-order behaviour of the transition (see next paragraph).

\begin{figure}[!h]
\begin{center}
\includegraphics[width=8.5cm,keepaspectratio]{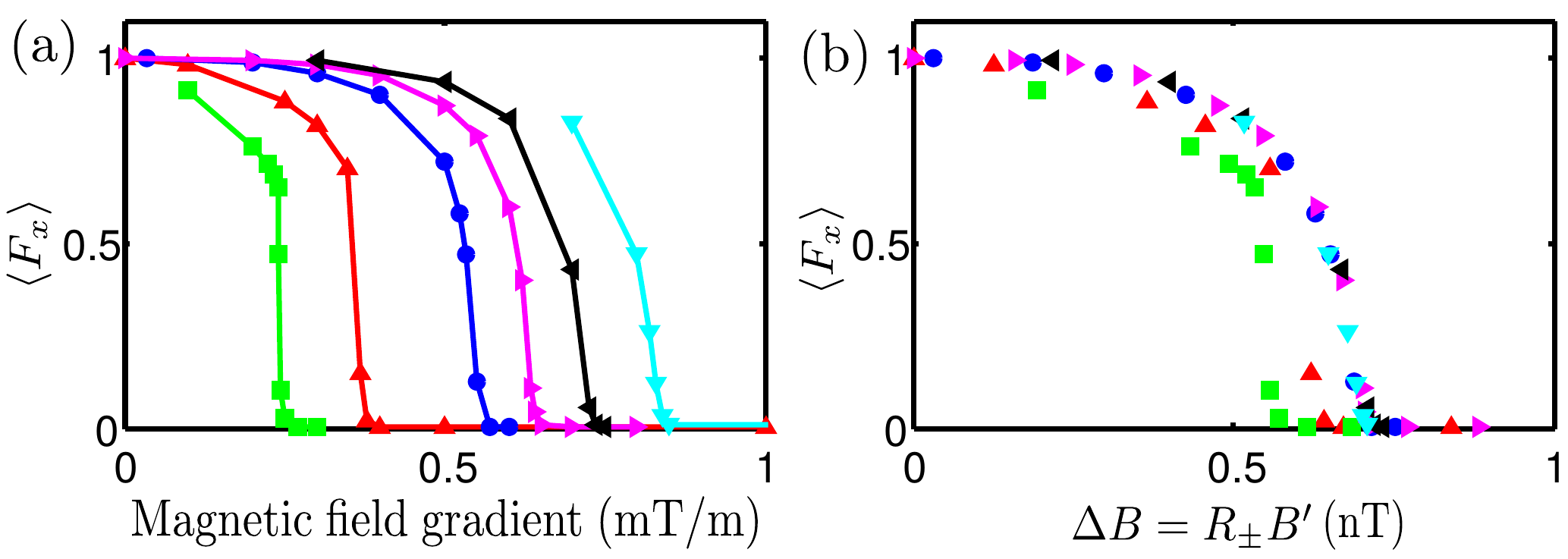}
\caption{Total magnetization along $x$ versus the gradient $B'$ (a) and the magnetic field difference across the condensate (b). Same legend as in Fig.~\ref{fig:res_sim_radius}.}
\label{fig:magnetization_vs_grad}
\end{center}
\end{figure}

In Fig.~\ref{fig:magnetization_vs_grad}~(a), we show the behaviour of the total magnetization along $x$, $\langle F_{x} \rangle \equiv \int \mathscr{F}_{x} (\mathbf{r}) \, d^{3}r$ \footnote{A local order parameter can be defined as the even part of $\mathscr{F}_{x}$ about $x$: $\Omega(x,y,z) \equiv \frac{1}{2} \left[ \mathscr{F}_{x}(x,y,z) + \mathscr{F}_{x}(-x,y,z) \right]$.}. We observe that, below a critical value of the gradient $B'_{c}$ of the order of 0.5~mT/m, the cloud is spontaneously magnetized along $x$, converging towards a fully polarized state for $B'=0$. The sudden change of magnetization when varying the gradient $B'$ is the signature of a phase transition, in particular the discontinuity at large atom number ($N \gtrsim 10^5$) would indicate a first order character for the transition. However, finite size effects cannot be easily excluded from the current simulation and further studies on the exact nature of the transition should be performed. We call weak gradient (WG) phase the phase for $B' < B'_{c}$, and strong gradient (SG) phase the one for $B' > B'_{c}$. The phase transition is also visible in Fig.~\ref{fig:res_sim_radius}: $\Rpm$ is independent of $\Delta B$ in the SG phase ($\Delta B \gtrsim 0.7$~nT for $N < 10^{5}$).

\begin{figure}[!h]
\begin{center}
\includegraphics[width=8.5cm,keepaspectratio]{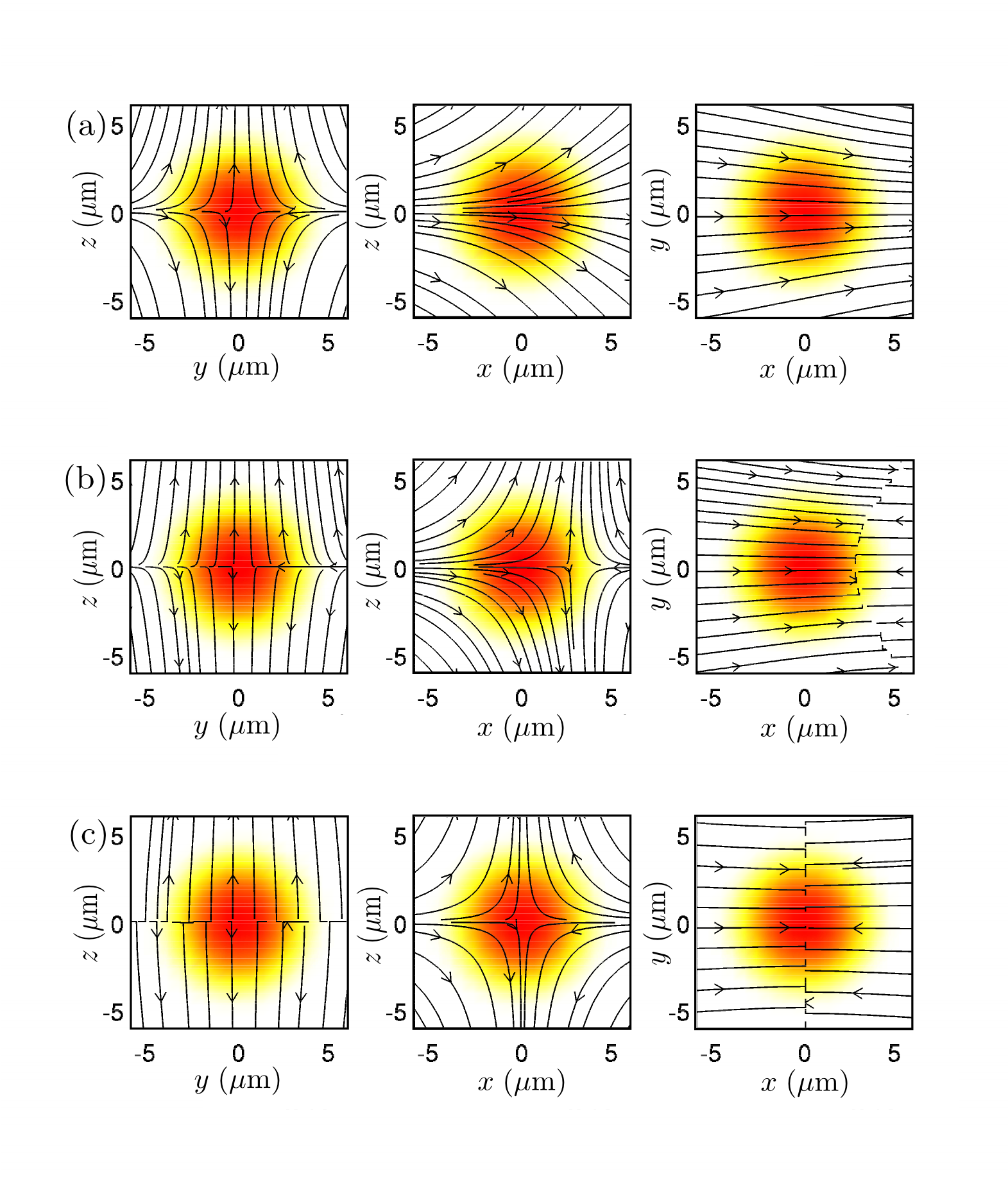}
\caption{(Color online) Spin streamlines (solid lines with arrow) and total density (color map) for $N=2 \times 10^{5}$. Slices in three orthogonal planes are presented, from left to right are slices in the $y-z$, $x-z$ and $x-y$ planes, respectively. (a) WG phase with $B'=0.4$~mT/m, (b) WG phase with $B'=0.535$~mT/m, and (c) SG phase with $B'=0.6$~mT/m.}
\label{fig:spin_streamlines}
\end{center}
\end{figure}

\noindent \textit{Broken symmetries} -- From the simulation results, we can study the WG and SG phases from a symmetry perspective. As already pointed out, $\mathrm{U(1)}_{\phi}$ and $\mathrm{SO(2)}_{F^{z} + L^{z}}$ are spontaneously broken as a result of the Bose/ferromagnetic transition. Thus, all continuous symmetries of $G$ are broken, and only discrete symmetries remain. In Fig.~\ref{fig:spin_streamlines}, we compare the spinor field streamlines, within slices of three orthogonal planes, for condensates in both phases. As pointed out earlier, here the prevailing axis $\hat{\mathbf{f}}$ is oriented along $x$.

In the WG phase [Figs.~\ref{fig:spin_streamlines}~(a) and (b)], the system is not invariant under $\mathsf{P}_{\hat{\mathbf{f}}}\mathsf{T}_{\hat{\mathbf{f}}}$, that is, the space-spin inversion along the axis $\hat{\mathbf{f}}$. Conversely in the SG phase [Fig.~\ref{fig:spin_streamlines}~(c)], the system is invariant under both $\mathsf{P}_{\hat{\mathbf{f}}}\mathsf{T}_{\hat{\mathbf{f}}}$ and $\mathsf{P}_{\hat{\mathbf{f}}_{\perp}}\mathsf{T}_{\hat{\mathbf{f}}_{\perp}}$, where $\hat{\mathbf{f}}_{\perp}$ is orthogonal to $\hat{\mathbf{f}}$ and lies in the $x-y$ plane.

The symmetries of the order parameter in a given phase constitute the \textit{isotropy group}, which is a subgroup of $G$. For the WG phase, the isotropy group is $H_{\rm WG} = \mathcal{PT}_{\hat{\mathbf{f}}} \times \mathcal{PT}_{z}$, where $\mathcal{PT}_{\hat{\mathbf{f}}} \equiv \left\{ \mathds{1}, \mathsf{P}_{\hat{\mathbf{f}}}\mathsf{T}_{\hat{\mathbf{f}}} \right\}$, whereas for the SG phase it is $H_{\rm SG} = \mathcal{PT}_{\hat{\mathbf{f}}} \times \mathcal{PT}_{\hat{\mathbf{f}}_{\perp}} \times \mathcal{PT}_{z}$.  

\noindent \textit{Order parameter space} -- Many properties of a given phase can be understood from the broken symmetries \cite{mermin1979}, mathematically defined as the quotient of the overall symmetry group by the isotropy group $H$ of the phase: $R=G/H$, called the \textit{order parameter space}. If $\Psi_{0}$ is an arbitrary order parameter of a given phase, called the \textit{standard order parameter}, then the order parameter manifold $\mathcal{M}$ of the phase results from the action of $R$ onto $\Psi_{0}$, that is, $\mathcal{M} = R\Psi_{0}$ \cite{sethna1992}. It results that many crucial properties of a phase, fully defined by the manifold $\mathcal{M}$, are embedded in the order parameter space $R$. This is in particular true for the topological properties.

The order parameter space of the WG phase is \footnote{We use the fact that the group $G$ can be factorized as $G = \mathrm{U(1)}_{\phi} \times \mathrm{SO(2)}_{F^{z} + L^{z}} \times \mathcal{PT}_{\hat{\mathbf{f}}/\hat{\mathbf{f}}_{\perp}} \times
\mathcal{PT}_{z}$.}
\begin{equation}
R_{\rm WG} = G/H_{\rm WG} = \mathrm{U(1)}_{\phi} \times \mathrm{SO(2)}_{F^{z} + L^{z}}.
\end{equation} 
From a topological perspective, both $\mathrm{U(1)}_{\phi}$ and $\mathrm{SO(2)}_{F^{z} + L^{z}}$ have the topology of a circle $\mathds{S}^{1}$, and the order parameter space is a \textit{torus}: $R_{\rm WG} \simeq \mathds{S}^{1} \times \mathds{S}^{1} = \mathds{T}^{2}$. For the SG phase, the order parameter space is
\begin{equation}
R_{\rm SG} = \mathrm{U(1)}_{\phi} \times \left( \mathrm{SO(2)}_{F^{z} + L^{z}} / \mathcal{PT}_{\hat{\mathbf{f}}} \right).
\end{equation} 
Let $\theta_{\hat{\mathbf{f}}}$ be the angle between the vector $\hat{\mathbf{f}}$ and the $x$ axis, then the quotient by the group $\mathcal{PT}_{\hat{\mathbf{f}}}$ identifies the elements of $\mathrm{SO(2)}_{F^{z} + L^{z}}$ according to the equivalence relation $\sim$, defined as for all $\theta, \theta' \in \mathds{R}$,
\begin{equation}
R_{z}(\theta) \sim R_{z}(\theta') \Leftrightarrow \left( \theta' = \theta \; \mathrm{or} \; \theta' = 2 \theta_{\hat{\mathbf{f}}} - \theta \right) \; \mathrm{mod} \; 2 \pi.
\end{equation}
As a consequence, the circle associated to $\mathrm{SO(2)}_{F^{z} + L^{z}}$ becomes a closed line, topologically equivalent to a closed interval: $\mathrm{SO(2)}_{F^{z} + L^{z}} / \mathcal{PT}_{\hat{\mathbf{f}}} \simeq \mathds{I}$, and the order parameter space of the SG phase is a \textit{cylinder}: $R_{\rm SG} \simeq \mathds{S}^{1} \times \mathds{I}$.

The knowledge of the order parameter space topology provides information on topological defects, in particular their stability. For both phases, the order parameter space is connected and therefore domain walls are unstable. However, these spaces are not simply connected, and thus stable 1D defects such as vortices can form. In the SG phase the fundamental group of $R_{\rm SG}$ is $\pi_{1} \left( \mathds{S}^{1} \times \mathds{I} \right) = \mathds{Z}$, and vortices are classified by a single winding number. Whereas for the WG phase, $\pi_{1} \left( \mathds{T}^{2} \right) = \mathds{Z} \times \mathds{Z}$ and the vortices are characterized by pairs of winding numbers, one related to the superfluid phase and the other to the magnetization. Moreover, higher order homotopy groups are trivial, and in particular point-like topological defects are unstable.

\noindent \textit{Conclusion and outlook} -- We have studied the ground-state properties of a ferromagnetic $F=1$ spinor condensate in the presence of a magnetic field gradient, a configuration that breaks both time-reversal ${\cal T}$ and parity ${\cal P}$ symmetries, but preserves the combined ${\cal PT}$ symmetry. Simulation reveals a phase transition that spontaneously breaks also this ${\cal PT}$ symmetry for weak gradient strength. Distinct topological defects are predicted in the weak- and strong-gradient phases. The fact that the polarization of the WG phase, parallel to $\hat{\mathbf{f}}$, is free to precess about the $z$ axis while protected by a phase transition suggests an attractive system for coherent field sensing. In contrast to other atomic field sensors \cite{BudkerNP2007,VengalattorePRL2007, ShahNP2007, KoschorreckAPL2011,SmithJPB2011,SewellPRL2012, BehboodAPL2013}, gradient-induced dephasing may be frozen out by the ferromagnetic interaction. This possibility motivates the study of the dynamics of this system under a combined gradient and bias field.

\noindent \textit{Acknowledgements} -- We thank Bruno Julia D{\'i}az, Artur Polls, and Luca Tagliacozzo for discussions. The work was supported by the Spanish MINECO projects MAGO (Ref. FIS2011-23520) and EPEC (FIS2014-62181-EXP), European Research Council project AQUMET, FET Proactive project QUIC and Fundaci\'{o} Privada CELLEX.


%

\end{document}